# Direct observation of the proliferation of ferroelectric loop domains and vortex-antivortex pairs

S. C. Chae[1], N. Lee[1], Y. Horibe[1], M. Tanimura[2], S. Mori[3], B. Gao[1], S. Carr[1†], and S.-W. Cheong[1*]

[1]Rutgers Center for Emergent Materials, Rutgers, The State University of New Jersey, Piscataway, NJ 08854

[2]Research Department, NISSAN ARC Ltd., Yokosuka, Kanagawa 237-0061, Japan

[3]Department of Materials Science, Osaka Prefecture University 1-1, Sakai, Osaka 599-8531, Japan, and JST, CREST, 1-1, Sakai, Osaka 599-8531, Japan.

*e-mail: sangc@physics.rutgers.edu

†current address: Columbia College, Columbia University

We discovered "stripe" patterns of trimerization-ferroelectric domains in hexagonal REMnO$_3$ (RE=Ho, ---, Lu) crystals (grown below ferroelectric transition temperatures ($T_c$), reaching up to 1435 $^o$C), in contrast with the vortex patterns in YMnO$_3$. These stripe patterns roughen with the appearance of numerous loop domains through thermal annealing just below $T_c$, but the stripe domain patterns turn to vortex-antivortex domain patterns through a freezing process when crystals cross $T_c$ even though the phase transition appears not to be Kosterlitz-Thouless-type. The experimental systematics are compared with the results of our six-state clock model simulation and also the Kibble-Zurek Mechanism for trapped topological defects.

The subtle phase transitions straddling the boundary between long-range-order and disorder have attracted significant attention due to the fundamental science and the technological perspective [1-5]. For example, the ground state of two-dimensional (2D) spins with planar continuous degrees of freedom undergoes only quasi-long-range-order with spin-spin correlations falling off algebraically in space, and becomes a high-temperature disorder state with exponentially-decaying correlations through the so-called Kosterlitz-Thouless (KT) transition by the proliferation of unbound topological vortices [6-8]. Not only the KT transition in 2D planar spins but also various phenomena such as 2D melting and roughening transition at surface are associated with the emergence of a topological order, resulting from the binding of topological defects [9, 10]. Even though the ordering issue of 2D condensed matters is a time-honored topic, the topological ordering process in "large-scale real space" has little investigated experimentally. Furthermore, it is a profound question how the topological KT order is influenced by the $3^{rd}$-dimensional coupling and quenched disorder that exists often in real systems.

Layered hexagonal $YMnO_3$ is an improper ferroelectric where the size mismatch between Y and Mn-O layers induces a trimerization-type structural phase transition, and this structural transition leads to three antiphase domains ($\alpha$, $\beta$, $\gamma$), each of which can support two directions (+,-) of ferroelectric polarization [11-13]. The antiphase and ferroelectric domains of $YMnO_3$ meet in cloverleaf arrangements that cycle through all six domain configurations. Occurring in pairs, the cloverleaves can be viewed as vortices and antivortices, in which the cycle of domain configurations is reversed [14]. Large-scale arrangements of topological vortices and antivortices in $YMnO_3$ reveal intriguing real space domain patterns with mathematically simplicity, which can be analyzed with graph theory [15]. The six possible characteristics of domains, combined with the layered structure of hexagonal $YMnO_3$, suggest the analogy between the 2D six-state clock model and the physics of $YMnO_3$. In the six-state clock model, it has been claimed the presence of three phases; long-range ordered (LRO) phase, intermediate KT phase, and high-temperature disordered phase [16]. However, vortex-antivortex domain patterns have been observed in $YMnO_3$ at room temperature far below structural transition temperature, which suggests the KT phase, rather than a LRO phase, as the ground state.

Herein, in order to unveil the origin of this inconsistency, we have studied the systematics of domain configurations in a series of hexagonal REMnO$_3$ (RE=rare earths) crystals grown with a flux method, and also the thermal evolution of the domain configurations. In order to reveal domain patterns, thin-plate-like crystals with optically flat surfaces were etched chemically in phosphoric acid at ~130 °C. The domain patterns of chemically-etched crystals were investigated using optical microscopy (OM), transmission electron microscopy (TEM) and atomic force microscopy (AFM). Note that preferential chemical etching of surface areas with upward polarization enables the observation of ferroelectric domain patterns on a crystal surface using OM or AFM. (see the detailed experimental method in the supplementary.)

Unlike vortex-antivortex domain patterns in YMnO$_3$, stripe domain patterns with large downward (-) polarization domains are discovered in most of REMnO$_3$ crystals. The OM image of a stripe domain pattern on one entire surface of an ErMnO$_3$ crystal is shown in Fig. 1(a). These stripe lines in the OM image are identified as narrow trenches with the depth of ~500 nm as reveled in AFM scans (see the inset of Fig. 1(a)), and tend to be along the [110] direction (the hexagonal *P6$_3$cm* notation). These trenched lines correspond to narrow upward (+) polarization domains [15]. These stripe domain patterns are distinct from the topological vortex-antivortex domain pattern with small domains in YMnO$_3$, shown in Fig. 1(b). In the vortex-antivortex domain pattern in YMnO$_3$, a vortex consists of six trimerization antiphase ($\alpha$, $\beta$, $\gamma$) and ferroelectric (+,-) domains merging at the center of the vortex, and is paired with an antivortex (or antivortices) with the opposite vorticity in terms of structural antiphase and ferroelectric relationship as shown in the inset of Fig. 1(b) [15]. On the other hand, the stripe domain pattern in Fig. 1(a) spans the entire crystal surface, and we have, in fact, observed only these stripe domain patterns without any hint of the presence of vortices in all REMnO$_3$ (RE=Ho, Er, Tm, Yb, Lu, but not Y) crystals. We also note that the stripe domain patterns with large downward-polarization domains appear to be consistent with an LRO phase, rather than the topological KT phase. As discussed below, this striking difference between stripe and vortex-antivortex domain patterns depends entirely on whether the crystal growth temperature is above the trimerization-structural transition temperature ($T_c$) or not.

In order to explore the thermal evolution of these stripe domain patterns, we cooled down crystals from various annealing temperatures to room temperature and then etched them chemically. Figure 2(a) displays the AFM image of a chemically-etched $ErMnO_3$ surface after cooling it down fast from 1120 °C. Even though the equilibrating temperature is very high, the LRO stripe domain pattern changes little. However, when the temperature is raised by only 20 °C, the pattern changes significantly, and exhibits highly-curved lines with the appearance of many curved closed loops, as shown in Fig. 2(b). Dark stripe lines, conserved robustly up to 1120 °C, start to wiggle heavily at 1140 °C, but never cross to each other, i.e., there is no hint of the presence of vortices in the entire crystal surface (see the supplementary section 2 and Fig. S2.). The more-or-less straight parts of dark stripe lines are indicated with white dashed lines in the upper region of Fig. 2(b). TEM image of closed loop domains is shown in Fig. 2(c), and the corresponding possible schematic is shown in Fig. 2(d). Therefore, when the system approaches $T_c$ from below, thermal fluctuations induce roughening of stripes domain walls and the appearance of a large number of loop domains.

When the annealing temperature is further raised by 30 °C up to 1170 °C, a complicated pattern was observed after chemical etching as shown in Fig. 2(e). This pattern, in fact, shows the crossing of lines and the presence of a large number of vortices. This vortex-antivortex domain pattern formation is more evident when the annealing temperature was raised to 1200 °C, as shown in Fig. 2(f). We have determined the characteristic temperatures for all $REMnO_3$ (RE=Ho, Er, Tm, Yb, Lu) at which stripe domain patterns turn into vortex-antivortex domain patterns after the annealing/cooling experiments. The obtained characteristic temperatures are plotted in the inset of Fig. 2(f), and the reported $T_c$ of $YMnO_3$ is also plotted in the inset [17]. The rough linear dependence in the inset strongly suggests that the characteristic temperatures, indeed, the trimerization-structural $T_c$ of $REMnO_3$. Note that $T_c$'s of $REMnO_3$ (RE=Ho, Er, Tm, Yb, Lu) have never been reliably determined because of the very high temperature nature, and $T_c$ drastically increases with decreasing RE size, which is consistent the notion that the structural transition is induced by the mismatch between small RE layers and large Mn-O layers in the $REMnO_3$ structure. Note that $REMnO_3$ crystals were grown by slow cooling of the materials with $Bi_2O_3$ flux in the temperature range of 1200 °C and 950 °C, but the real growth through nucleation occurs probably slightly above 950 °C. Thus, $YMnO_3$ crystals are likely grown above $T_c$, but other REMnO$_3$ crystals below $T_c$. Therefore, it appears that stripe domain patterns form when the crystal growth temperature is below $T_c$ while vortex-antivortex domain patterns are realized when crystals are exposed to temperatures above $T_c$.

Interestingly, we found that once vortex-antivortex domain patterns, spanning the entire crystal surface (see the supplementary section 3 and Fig. S3), form by crossing $T_c$, they are conserved with various thermal treatments, but the domain size of vortex-antivortex domain patterns or the distance between vortices and antivortices can vary in a systematic manner. In order to find out the thermal evolution of vortex-antivortex domain patterns and the domain growth kinetics, the cooling rate near $T_c$ was changed from 0.5 $^o$C/h to 300 $^o$C/h. In addition, we cooled one specimen from 1220 $^o$C to 677 $^o$C with cooling rate of 5 $^o$C/h, followed by quenching to room temperature. Figures 3(a), (b), (c) and (d) show the AFM images of etched ErMnO$_3$ crystals with the above thermal treatments. With the large variation of cooling rate from 0.5 $^o$C/h to 300 $^o$C/h, vortex-antivortex domain patterns remain intact, but the domain size of vortex-antivortex domain patterns changes systematically.

We emphasize that the vortex-antivortex domain patterns in Figs. 3(b) and (c) are basically identical, indicating that the cooling rate below 677 $^o$C does not influence the domain patterns. This is an important result for the origin of the mysterious second transition of REMnO$_3$ near 600 $^o$C reported in many early publications [12, 17-24]. This second transition at ~600 $^o$C was identified as the ferroelectric transition from centrosymmetric *P6$_3$/mmc* to low-temperature polar *P6$_3$cm* structures via an intermediate *P6$_c$/mcm* structure [18, 19] whereas other results argued for the nonexistence of intermediate *P6$_3$/mcm* state, but the presence of an isosymmetric phase transition with Y-O hybridization [12, 17, 20, 21, 23]. We, in fact, compared directly vortex-antivortex domain patterns at room temperature and 730 $^o$C from TEM dark-field experiments using the 1 $\bar{3}$ 1 diffraction spot as shown in Figs. 4(a) and (b). Basically there is little difference between two vortex-antivortex domain patterns, which, combined with no difference in vortex-antivortex domain patterns on cooling rate across the second transition temperature, are consistent with the possibility of an isosymmetric change at the second transition if it exists.

The analysis of the evolution of domain size of vortex-antivortex domain patterns with varying cooling rate demonstrates slow growth kinetics associated with the topological vortex-antivortex domain patterns. The average distance of vortex-antivortex pairs vs. inverse cooling rate, $t$, is plotted in Fig. 4(c). The cooling rate dependence of the average vortex-antivortex pair distance, $D$, can be described by $D \propto t^n$ with $n$=0.23 (See the supplementary section 4 and Fig. S4). This value of $n$≈1/4 is rather different from the typical parabolic domain growth value of $n$=1/2 [25, 26]. This slow growth kinetics with vortex-antivortex domain patterns seem consistent with the Kibble-Zurek Mechanism for spontaneous trapping of topological defects in a system undergoing a continuous phase transition [27, 28]. For the average vortex-antivortex pair distance $\propto t^n$ with $n$≈1/4, the density of vortex should vary like $t^{1/2}$, which is precisely the prediction of the Kibble-Zurek mechanism with mean field critical exponents [28]. We also note that chemical/structural-defects-induced pinning may play an important role for the variation of the vortex-antivortex distance with cooling rate. In fact, strong pinning tendency of domain patterns has been observed in many of our results. For example, the domain patterns in Fig. 2(a) and Fig. 4(b) change little with expected large thermal fluctuations at temperatures such as 1120 °C and 730 °C. In addition, we have directly observed vortex-antivortex domain patterns pinned by large-scale surface defects, as shown in Fig. 4(d). Strong pinning, slow kinetics, and the different topology between vortex-antivortex and stripe domain patterns induce probably an "astronomical" time scale for the conversion of vortex-antivortex domain patterns to stripe domain patterns, even though the true ground state corresponds to stripe domain patterns.

The concept of topological KT order with binding vortices and antivortices was originally developed for the quasi-long-range-order in 2D XY systems. In the KT phase, algebraically decaying correlation can accompany topological defects, i.e., vortices. As the order parameter possesses a finite value, q, representing the number of evenly-spaced possible order orientations, there exists a long-range ordered phase at low temperatures when q is small. Convincing evidence has been amassed, based both on qualitative and quantitative considerations, to indicate that for large enough q (thought to be 5≤q<8), there exists an intermediate phase similar to the KT phase [16]. The 2D q-state clock model with q≥8 precisely reproduces the KT transition [29, 30]. In the case of the 2D six-state clock model, q=6, the topological order exists only in an intermediate temperature range, and the ground state is a LRO phase (see the supplementary section 5 and Fig. S5.). Any $3^{rd}$ dimensional coupling will destabilize the intermediate KT phase in the six-state clock model. Therefore, one expects theoretically that a LRO phase will be the ground state of hexagonal REMnO$_3$ with six degrees of freedom, and the intermediate KT order may or may not exist, depending on the strength of $3^{rd}$ directional coupling. Experimentally, our results indicate that the LRO ground state with stripe domain patterns can be observed, but only when crystals were grown below $T_c$. When crystals were grown from temperatures above $T_c$ or crystals with stripe domain patterns were exposed to a temperature above $T_c$, domain patterns are always vortex-antivortex domain patterns. In addition, we have observed convincing evidence for strong pinning and also slow kinetics associated with domain patterns. Furthermore, our results do not show any indication for two step transitions from a long-range order to a KT order to disorder. Therefore, it is convincing that there exists only one LRO transition associated with six degrees of freedom. However, in any real situations without astronomical-time-scale annealing, slow kinetic creation of topological defects, which is well manifested in the Kibble-Zurek mechanism, and their strong pinning induce KT-type vortex-antivortex domain patterns when crystals undergoes the structural transition at $T_c$. In other words, a KT order is only implicit in hexagonal REMnO$_3$ and should not occur as the ground state, but it is arrested by slow kinetic and strong pinning. This "arresting" scenario is fully consistent with our simulation of the 2D six-state clock model with different cooling rates (see the supplementary section 5 and Fig. S5.).

In summary, the true ground state with stripe domain patterns can be realized when hexagonal REMnO$_3$ (RE=Ho, Er, Tm, Yb, Lu) crystals are grown below $T_c$, and is consistent with the long-range-ordered ground state of the six-state clock model with a significant $3^{rd}$ dimensional coupling. When crystals cross $T_c$, domain patterns with topological vortices and antivortices are realized. These vortex domains can be enlarged with the time exponent of 1/4, which is much smaller than the typical domain growth exponent of 1/2. It is conceivable that the topological KT-type vortex-antivortex domain patterns eventually turn into long-range-ordered stripe domain patterns through astronomical-time-scale annealing. Therefore, the presence of stripe domain patterns in REMnO$_3$ crystals grown below $T_c$ is an experimental marvel, enabling the observation of what happens in the true ground state of REMnO$_3$.

Acknowledgement:

We thank D. Vanderbilt, W. Zurek and V. Zapf for critical reading of the manuscript and helpful discussions. This work was supported by National Science Foundation DMR-1104484.

**Figure legends**

**FIG 1.** (color online). Two distinct domain patterns of REMnO$_3$ crystals; stripe vs. vortex patterns. (a) Optical microscope image of a chemically-etched ErMnO$_3$ crystal surface. The inset shows the atomic force microscope (AFM) image of a stripe domain pattern after chemical etching. These domain patterns for an long-range ordered phase are also observed in HoMnO$_3$, TmMnO$_3$, YbMnO$_3$, LuMnO$_3$ crystals, as shown in Fig. S1 in Supplementary. (b) Optical microscope image of a chemically-etched YMnO$_3$ crystal surface. The inset shows the AFM image of an YMnO$_3$ surface after chemical etching, showing a vortex-antivortex domain pattern. Vortex and antivortex are distinguished by the arrangement of trimerization antiphase and ferroelectric domains with the opposite sense of rotation around a core. Note that the entire surfaces of all as-grown crystals REMnO$_3$ (RE=Ho, Er, Tm, Yb, Lu) exhibit no hint of the presence of any vortices or antivortices.

**FIG 2.** (color online). The thermal evolution of stripe domain patterns in ErMnO$_3$. (a), (b), (e) and (f), The AFM images of chemically-etched ErMnO$_3$ crystals annealed at 1120 $^o$C, 1140 $^o$C, 1170 $^o$C, and 1200 $^o$C, respectively. A stripe domain pattern with thermal roughening are evident in (b), but (e) and (f) display vortex-antivortex patterns. The inset displays $T_c$'s of REMnO$_3$ estimated from the formation temperature of vortex-antivortex domain patterns. $T_c$ of YMnO$_3$ is from ref. [17]. (c) and (d), TEM image of loop domains of ErMnO$_3$ quenched from 1140 $^o$C and a possible corresponding schematic, respectively. The loop domains observed in the TEM image are assigned with three antiphase domains ($\alpha+$, $\beta-$, $\gamma+$).

**FIG 3.** (color online). The evolution of vortex-antivortex domain patterns with varying cooling rate. The AFM images of chemically-etched ErMnO$_3$ crystals; (a) cooled from 1220 $^o$C to 890 $^o$C with rate of 0.5 $^o$C/h, followed by furnace cooling, (b) cooled from 1200 $^o$C to room temperature with rate of 5 $^o$C/h, (c) from 1200 $^o$C to 677 $^o$C with rate of 5 $^o$C/h, followed by quenching, and (d) from 1200 $^o$C to room temperature with rate of 300 $^o$C/h.

**FIG 4.** (color online). The pinning of vortex-antivortex domains. (a) and (b) TEM images of vortex-antivortex domain patterns of ErMnO$_3$ at room temperature and 730 $^o$C, respectively. (c) The cooling rate dependence of the average distance of vortex-antivortex pairs, exhibiting a power law dependence with the power of 0.23. (The open triangle is from Fig 3(c)). (d) The optical microscope and AFM (inset) images of a chemically-etched $ErMnO_3$ crystal surface, suggesting strong pinning of vortex-antivortex domains by surface defects.

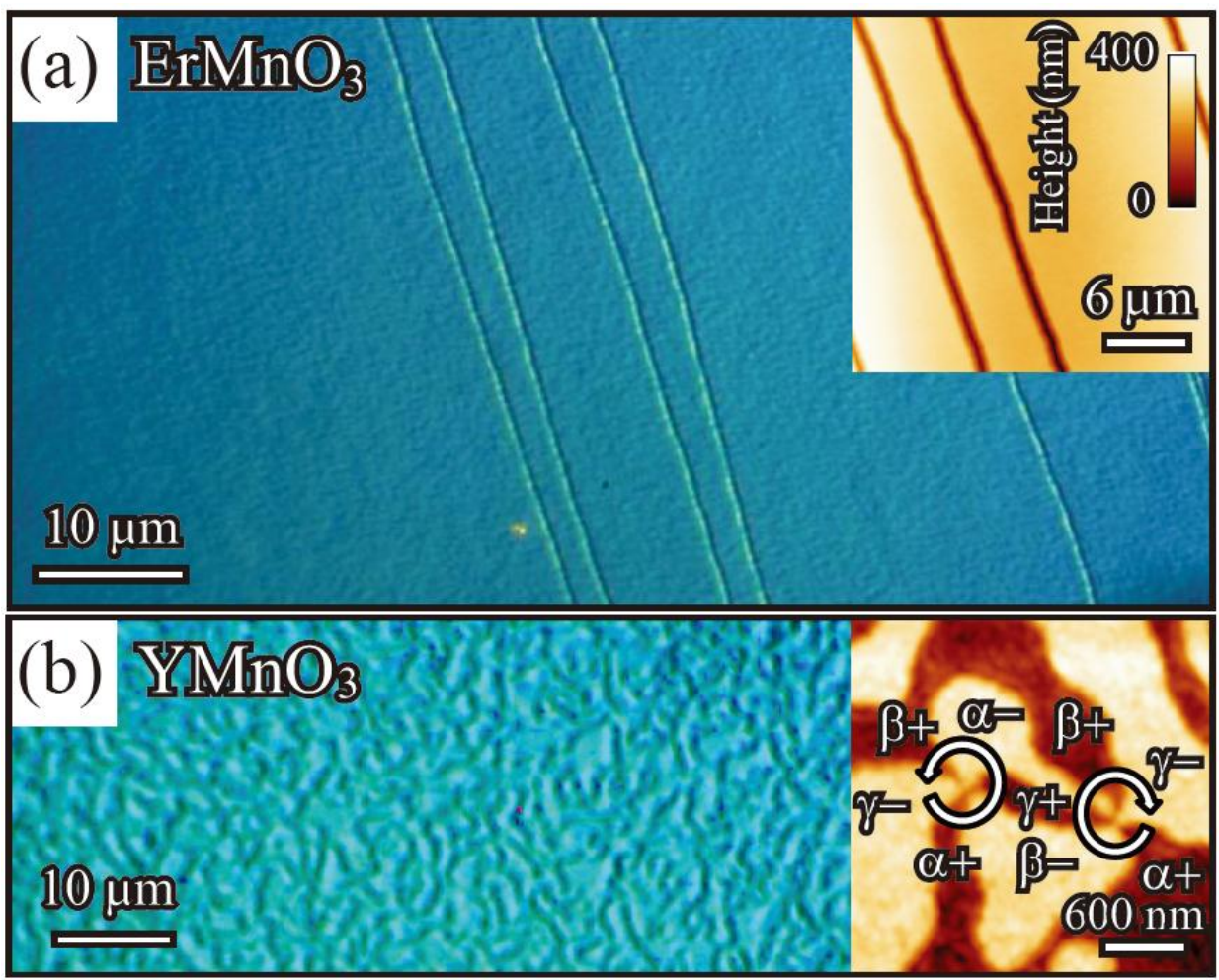

*Figure 1*

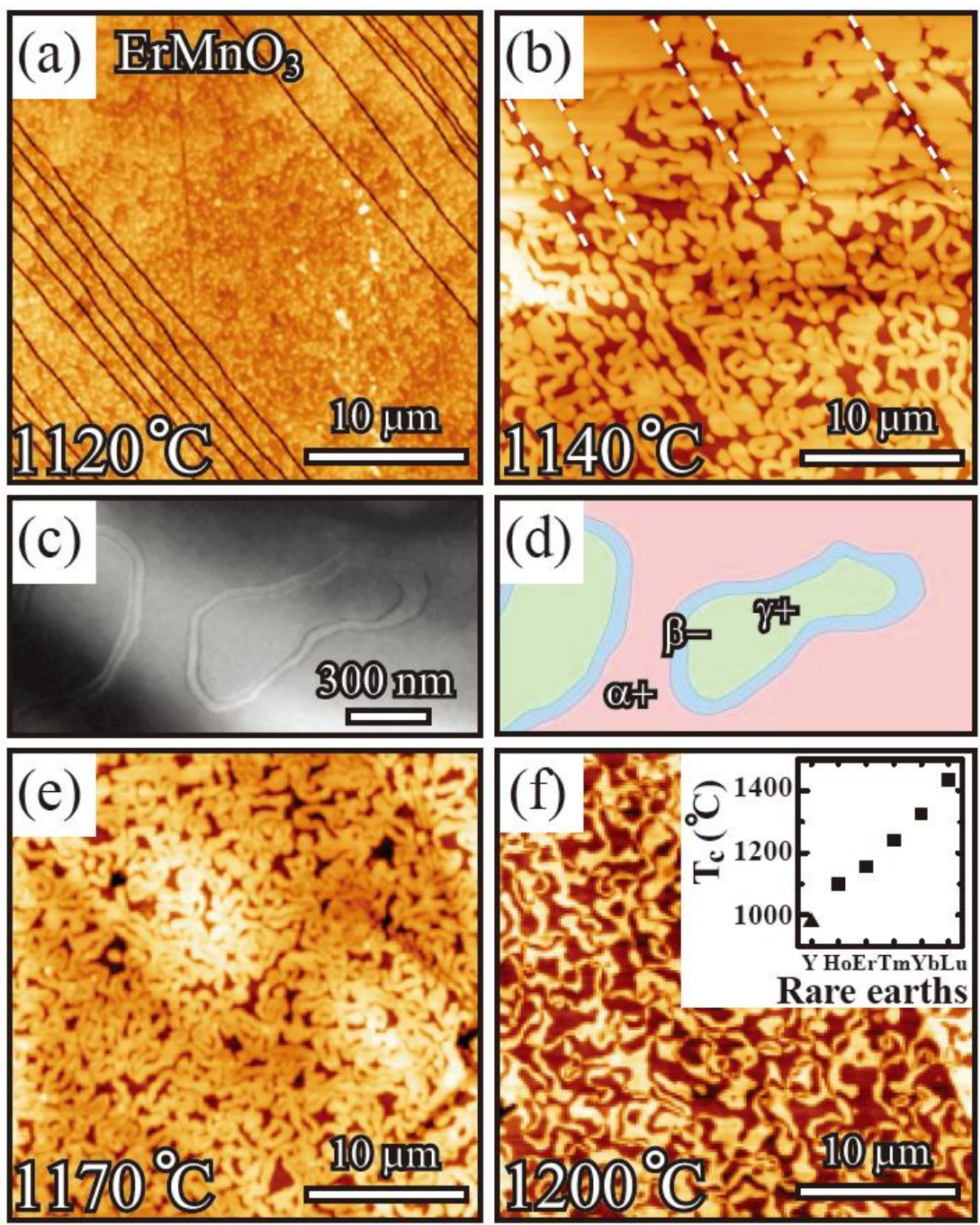

The figure panels contain the following labels:

(a) ErMnO₃ — 1120°C — 10 μm

(b) 1140°C — 10 μm

(c) 300 nm

(d) β— γ+ α+

(e) 1170°C — 10 μm

(f) 1200°C — 10 μm

Inset (f): $T_c$ (°C) — Y Ho Er Tm Yb Lu — Rare earths

*Figure 2*

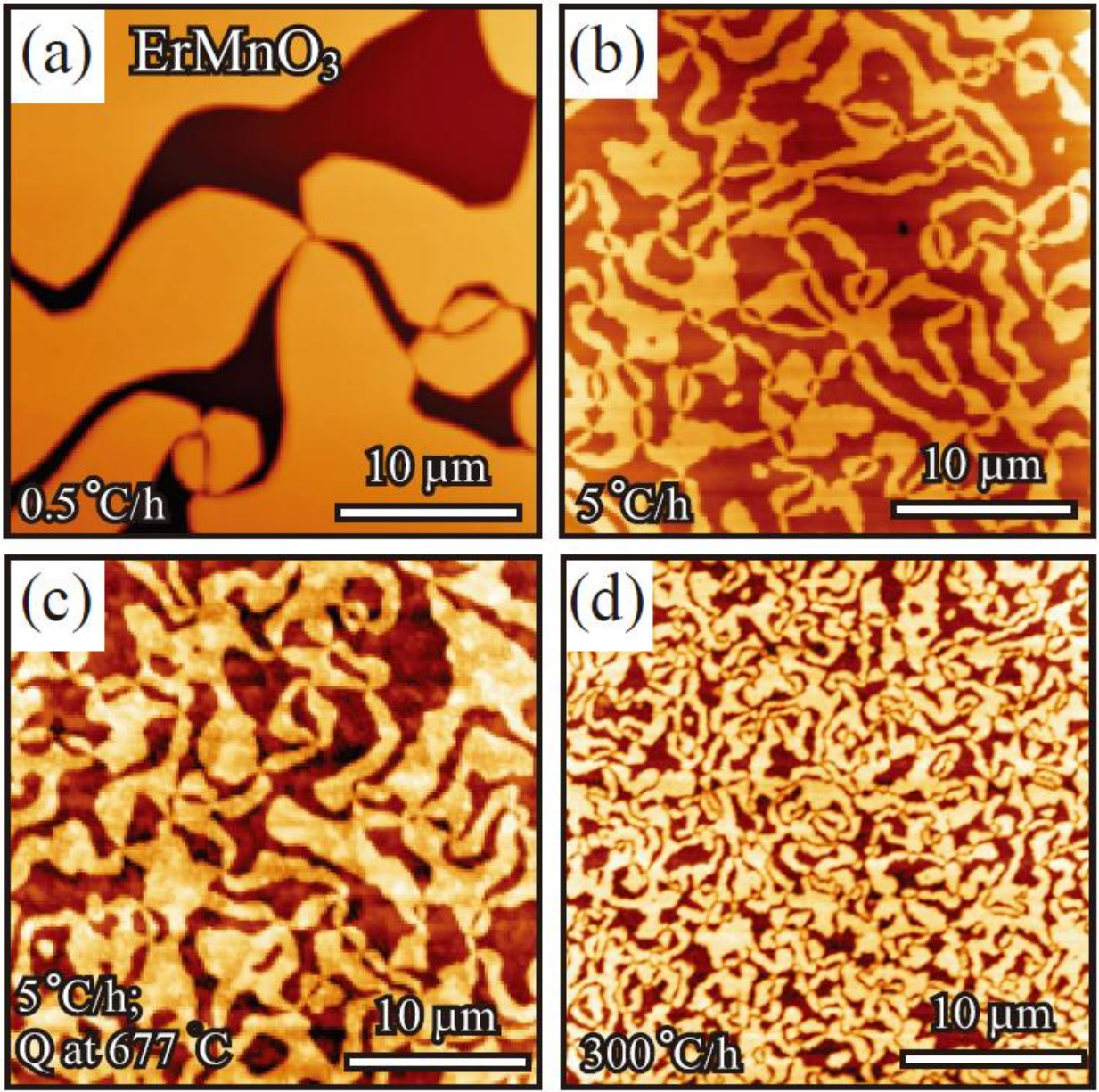

*Figure 3*

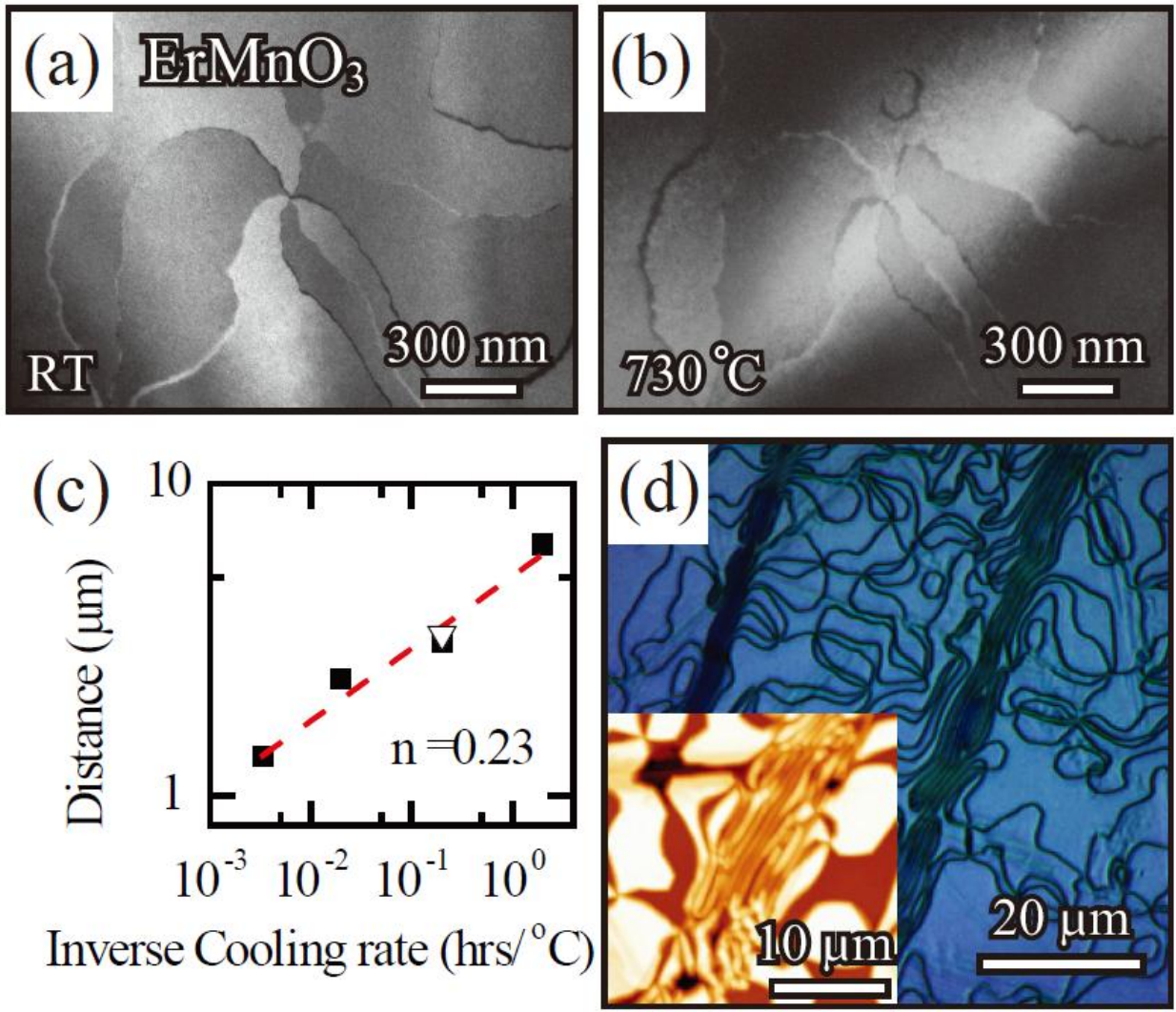

*Figure 4*

# Direct observation of the proliferation of ferroelectric loop domains and vortex-antivortex pairs

S. C. Chae, N. Lee, Y. Horibe, M. Tanimura, S. Mori, B. Gao, S. Carr, and S.-W. Cheong

Supplementary Information

## Experimental method

Thin-plate-like $REMnO_3$ single crystals with a few square millimeters in in-plane size were grown using a flux method with a mixture of 90 mol% of $Bi_2O_3$ and 10 mol% of $REMnO_3$ powders, which was slowly cooled from 1200 $^o$C to 950 $^o$C, followed by furnace cooling. To observe domain configurations, crystals with two wide natural facets normal to the crystallographic $c$ axis of hexagonal $REMnO_3$ ($P6_3cm$) were etched chemically in phosphoric acid for 30 minutes at 130 $^o$C. The transmission electron microscopy (TEM) observation was carried out using both JEOL-2010F and JEOL-2000FX TEM on $ErMnO_3$ single crystals at various temperatures. Specimens for dark-field TEM imaging experiments were prepared utilizing Ar ion milling at the liquid-nitrogen temperature. The atomic force microscopy (AFM) observation was carried out using a Nanoscope IIIA (Veeco) on the surfaces of chemically-etched $REMnO_3$ single crystals.

In order to observe how stripe domain patterns (see Section S1) evolve with varying temperature especially near $T_c$. Crystals were quenched from high temperature and then etched chemically. $T_c$ turns out to be impractically high in order to perform any in-situ real-space imaging experiments. Therefore, we have made domain pattern observations on $REMnO_3$ crystals quenched after equilibrating them at high temperatures near $T_c$ with the expectation that the high-temperature domain configurations do not change during quenching due to the lack of domain growth time. The results of our quenching experiments reveal that long-range ordered stripe domain patterns without vortices remain intact (except the appearance of small closed loop domains) as long as the equilibrating temperature is below $T_c$, but become immediately topological vortex-antivortex domain patterns as soon as the equilibrating temperature is above $T_c$.

**Section S1. Long-range ordered stripe domain patterns of h-REMnO$_3$.**

The stripe patterns of mutually-interlocked ferroelectric and structural antiphase domains, showing long-range ordered states, are observed in various h-REMnO$_3$ (RE=Ho, Tm, Yb, Lu) crystals grown with a flux method, as shown in Fig. S1. Utilizing different etching rates for different polarization directions, surface ferroelectric domain patterns are revealed by chemical etching. When the crystals are heated above, and then cooled below the ferroelectric-antiphase transition temperature, these stripe domain patterns turn into vortex-antivortex domain patterns, reflecting the topological Kosterlitz–Thouless (KT) phase.

**Section S2. A thermally-induced domain pattern with highly curved stripes and loop domains.**

When crystals with stripe domain patterns are annealed at a temperature just below the ferroelectric-antiphase transition temperature, stripe domains in long-range ordered stripe patterns become highly curved, and numerous loop domains appear, as shown in Fig. S2a. These stripes and loops never cross each other, i.e., they never show any hint of the presence of ferroelectric-antiphase vortices as shown in Figs. S2b and S2c. As the annealing temperature approaches the ferroelectric-antiphase transition temperature, loop domains appear and span the entire surface as shown in Fig. S2d. However, as soon as the annealing temperature is above the ferroelectric-antiphase transition temperature, a vortex-antivortex domain pattern with numerous ferroelectric-antiphase vortices and antivortices appears.

**Section S3. Large vortex-antivortex domains on the entire surface of a slowly-cooled ErMnO$_3$ crystal.**

Using the optical microscope, we have observed a pattern of large vortex-antivortex domains on the surface of an ErMnO$_3$ crystal, cooled with rate of 0.5 °C/h across the structural transition temperature (see Fig. S3). No trivial irregularity was observed in the pattern on the entire crystal surface, and this pattern can be considered as a snapshot of a kinetically-arrested KT phase.

**Section S4. The average domain size of vortex pattern.**

In the main text, the exponent $n$ for domain size growth was estimated by assuming that the average vortex-antivortex distance is similar with (or proportional to) the average size of domains. When domains are irregular, the conventional way to estimate the average domain size is checking the ratio between area and the total domain wall length [1]. With this method, we have estimated the average domain size for each cooling rate. We found that the "domain-size" growth with time shows also a power law dependence with the exponent of ~0.22 (see Fig. S4). This exponent is highly consistent with the value (0.23) from the average vortex distance. The consistency demonstrates beautifully the scale invariance of the domain patterns with different cooling rates.

**Section S5. Monte Carlo simulation of the six-state clock model with varying cooling rate.**

We have studied the dependence of the domain patterns of the 2D six-state clock model on inverse cooling rate using a Monte Carlo simulation [2]. The Hamiltonian for the six-state clock model is

$$H = -J \sum_{<i,j>} s_i \cdot s_j = -J \sum_{<i,j>} \cos(\theta_i - \theta_j),$$

where each order parameter, $s_i$, can make angles $\theta_i = \pi n_i / 3$ ($n_i$=1,...,6). The sum is over nearest neighbors on a hexagonal lattice with the coupling constant, $J > 0$. The simulation was performed on a hexagonal lattice of the 100 × 100 size with periodic boundary conditions. In order to investigate the quenching effect on the domain pattern, we changed the Monte Carlo Step (MCS) unit – the number of iterations – from 10 to 1000 during the simulation temperature was reduced in steps; The higher MCS unit was used for the simulation, the more iteration at fixed temperature was done to mimic the slow-cooling of crystal. Here the temperature was varied from 2.1 to 0.1 K with step of 0.2 K. Figure S4 shows the dependence of the simulation results on temperature and inverse cooling rate. Each column represents the simulated domain patterns for various temperatures for a fixed inverse cooling rate. Each row represents the simulated domain pattern at each temperature. For the slowest cooling rate (MCS=1000), the temperature evolution is consistent with the presence of two transitions: one between disorder and a KT order and the other between a KT order and long-range order. At the lowest temperature, the single domain pattern representing the true ground state with an long-range order for slow cooling rate evolves to a KT-order-like pattern with vortices and antivortices with increasing cooling rate. These simulation results are consistent with our experimental observation of various domain patterns with different cooling rates as well as different growth conditions.

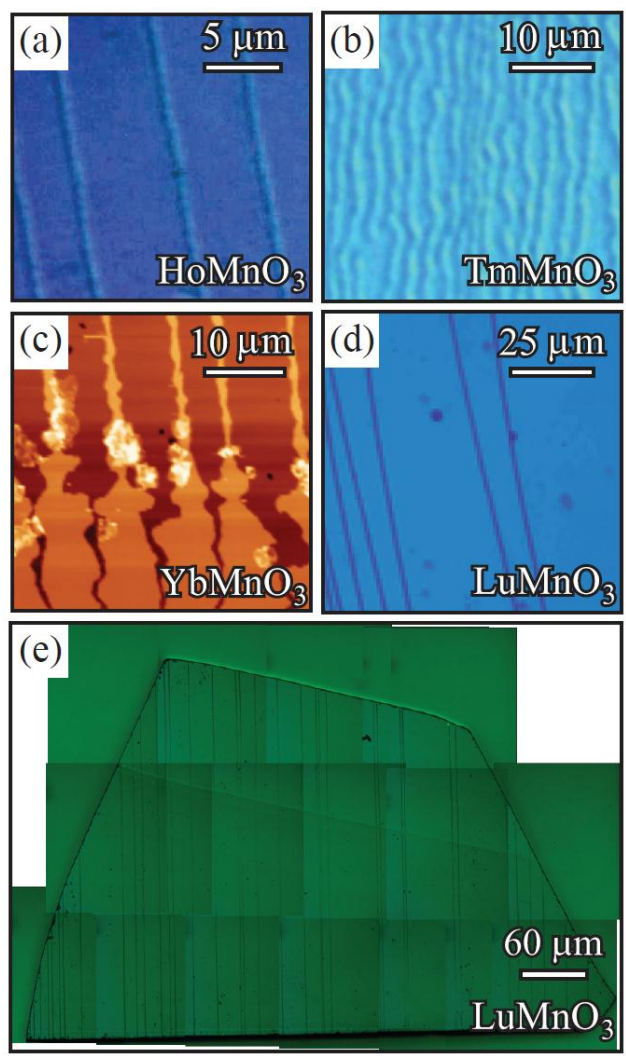

**Figure S1 | Ising-type stripe domain patterns of h-REMnO₃ (RE=Ho, Tm, Yb, Lu). a, b, and d,** Optical microscope images of HoMnO₃, TmMnO₃, and LuMnO₃ (0001) surfaces, respectively. **c,** Atomic force microscope image of an YbMnO₃ (0001) surface. The optical microscope images and atomic force microscope images were obtained on crystal surfaces etched chemically using phosphoric acid. **e,** the entire surface of a chemically-etched LuMnO₃ crystal without any evidence for vortex.

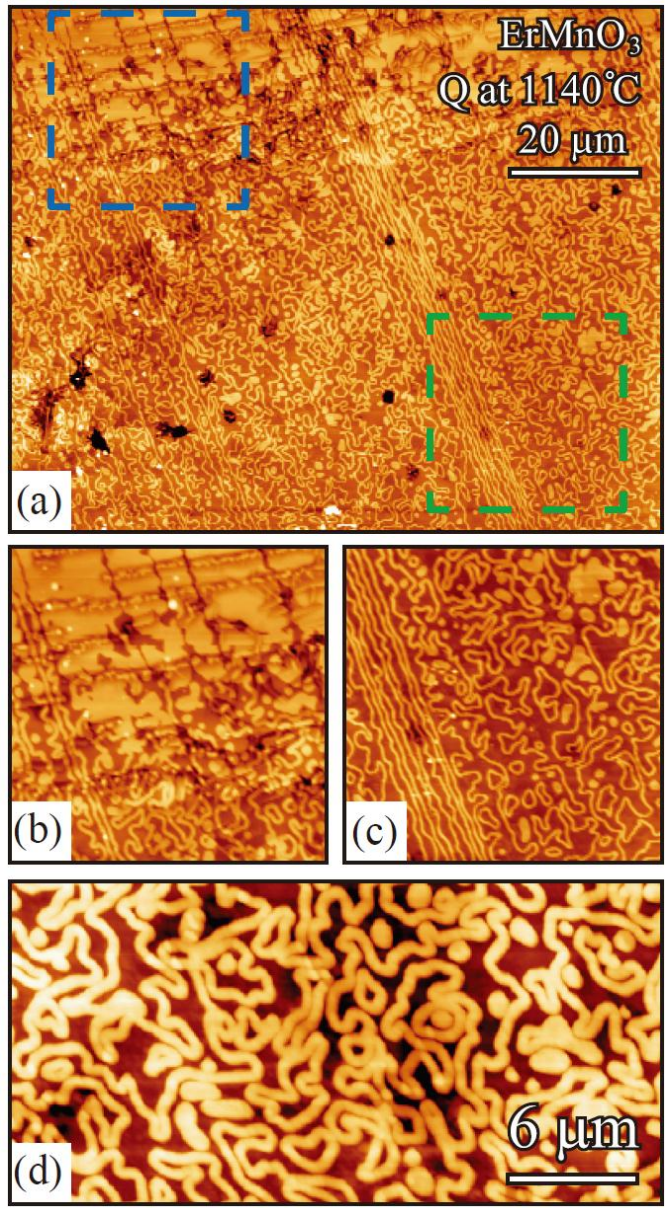

**Figure S2 | A thermally-induced stripe domain pattern with highly curved stripes and numerous closed loops. a,** Atomic force microscope image of the chemically-etched surface of an ErMnO$_3$ crystal quenched from 1140 $^o$C shows highly-curved stripe domains and also many closed loop domains. **b,** The enlarged image of the blue rectangle in Fig. S2a. **c,** The enlarged image of the green rectangle in Fig. S2a. Careful inspection reveals that narrow stripes and closed loop domains never cross to each other. **d.** High-resolution image of loop domains.

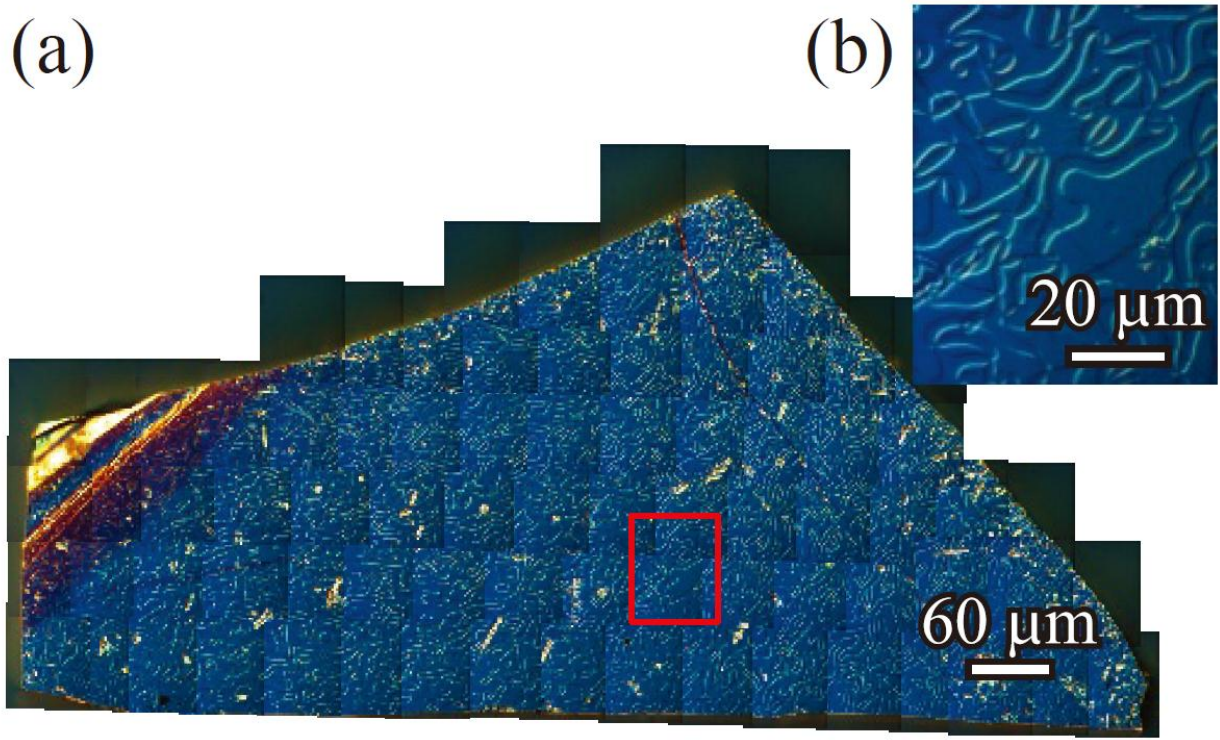

**Figure S3 | Vortex-antivortex domain pattern of ErMnO₃, cooled slowly from 1220 °C to 890 °C with rate of 0.5 °C/h. a,** The picture was collaged with a large number of optical microscope images taken on a crystal surface after chemical etching. **b,** The enlarged optical microscope image of the area depicted with the red rectangle in Fig. S3a.

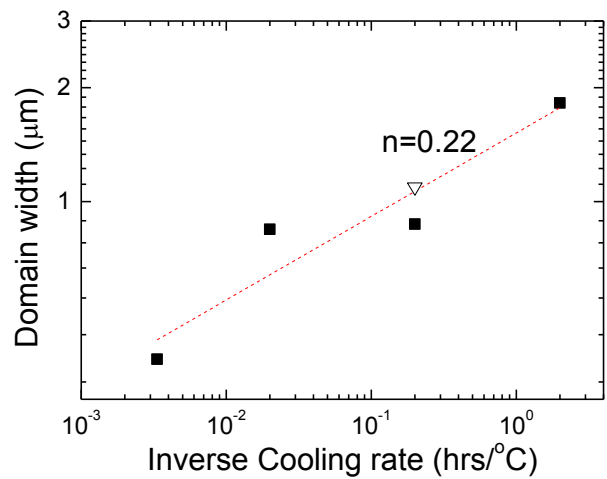

**Figure S4| The cooling rate dependence of the average domain size, exhibiting a power law dependence with the exponent of 0.22.**

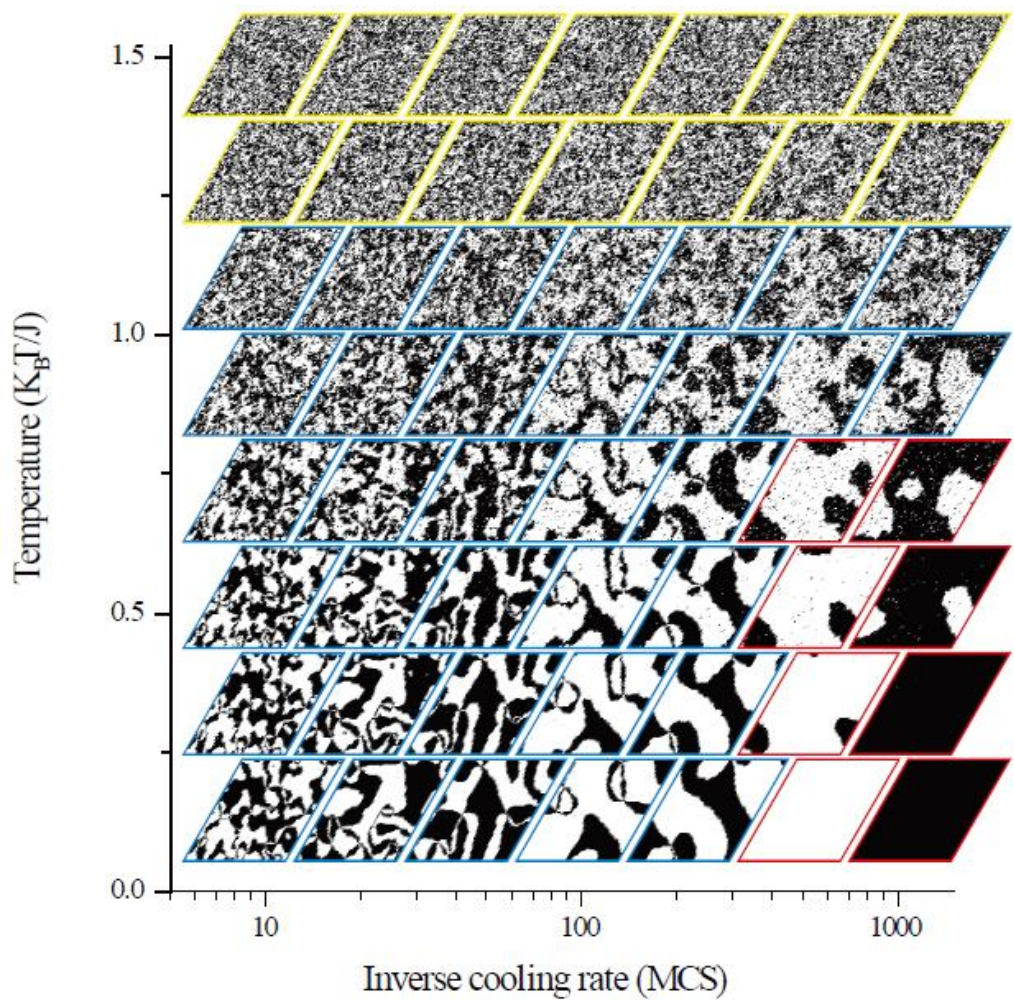

**Figure S5 | Monte Carlo simulation of the six-state clock model.** Domain patterns at various temperatures with various cooling rates are obtained from a Monte Carlo simulation for the six-state clock model with a hexagonal lattice. The yellow, blue and red colors of the frames of domain patterns correspond to disordered, vortex-antivortex domain and long-range ordered domain patterns, respectively.